\documentclass[a4paper,eqsecnum,english,a4paper,showpacs,amsmath,amssymb,twocolumn]{revtex4}
\usepackage[latin1]{inputenc}
\usepackage{float}
\usepackage{graphicx}
\usepackage{amssymb}

\makeatletter

\usepackage{dcolumn}

\usepackage{bm}

\usepackage{babel}
\makeatother
\begin{document}

\title{Collective excitations in Na$_2$IrO$_3$}

\author{Jun-ichi Igarashi$^{1}$ and Tatsuya Nagao$^{2}$}

\affiliation{
 $^{1}$Faculty of Science, Ibaraki University, Mito, Ibaraki 310-8512,
Japan\\
$^{2}$Faculty of Engineering, Gunma University, Kiryu, Gunma 376-8515,
Japan
}

\date{\today}

\begin{abstract}
We study the collective excitations of Na$_2$IrO$_3$ 
in an itinerant electron approach. 
We consider 
a multi-orbital tight-binding model 
with the electron transfer between the Ir $5d$ states 
mediated via oxygen $2p$ states and the direct $d$-$d$ transfer
on a honeycomb lattice. 
The one-electron energy as well as the ground
state energy are investigated within the Hartree-Fock approximation.
When the direct $d$-$d$ transfer is weak, we obtain nearly flat energy
bands due to the formation of quasimolecular orbitals, and the ground
state exhibits the zigzag spin order. 
The evaluation of the density-density correlation function within the 
random phase approximation shows that
the collective excitations emerge as bound states. 
For an appropriate value of the direct $d$-$d$ transfer,
some of them are concentrated in the energy region 
$\omega <$ 50 meV
(magnetic excitations) while the others lie in the energy region 
$\omega >$ 350 meV (excitonic excitations). This behaviour is consistent
with the resonant inelastic x-ray scattering 
spectra.
We also show that the larger values of the direct $d$-$d$ transfer
are unfavourable in order to explain the observed aspects of Na$_2$IrO$_3$
such as the ordering pattern of the ground state
and the excitation spectrum.
These findings may indicate that the direct $d$-$d$ transfer is 
suppressed by the structural distortions
in the view of excitation spectroscopy, 
as having been pointed
out in the \emph{ab initio} calculation. 
\end{abstract}


\maketitle

\section{Introduction\label{sect.1}}
The physics of $5d$-based iridates has recently attracted much attention,
since the competition between the large spin-orbit interaction (SOI) and 
the Coulomb interaction makes their physical properties quite different 
from those of the $3d$ transition metal compounds.
Novel phases such as the topological insulator, the Weyl semimetal,
and spin liquid have been explored extensively in these materials
\cite{Krempa2014,Rau2015}.
In particular, these research activities may have been accelerated
by the waves of new discovery supplied by some representative materials.

One such example is Sr$_2$IrO$_4$, which shows the antiferromagnetism with 
the spin-orbit coupled \emph{isospin} $j_{\rm eff}=1/2$.
The low-lying excitations have been detected by resonant inelastic 
x-ray scattering (RIXS), where the magnon peak exists at $\omega<0.2$ eV 
and the exciton peaks emerge around $\omega\sim 0.5$ eV 
with substantial weights as a function of energy loss $\omega$ 
\cite{Ishii2011,J.Kim2012,Crawford1994,Cao1998,Moon2006}. 
On the localized electron picture, the Heisenberg-type spin Hamiltonian 
has been derived by the strong-coupling theory\cite{Jackeli2009,Kim2012,Katukuri2012}. 
The spin Hamiltonian seems to provide
a good description for the observed magnetic excitations.
Recently it has been predicted that the magnon is split into two modes 
due to the interplay between Hund's coupling and the SOI 
\cite{Igarashi2013-2,Igarashi2014-1,Vladimirov2014,Vladimirov2015}.
Such band splitting has now been
confirmed by the magnetic critical scattering experiment \cite{Vale2015}
and RIXS \cite{J.Kim2014}.

As regards the itinerant electron picture, the band structure calculation 
has been carried out within the density functional theory (DFT) \cite{Kim2008}. 
It provides an insulating antiferromagnetic ground state.
Recently the collective excitations have been investigated by introducing
a multi-orbital tight-binding model by present authors\cite{Igarashi2014-1,Igarashi2014-2,Com3}.
The density-density correlation function 
has been investigated within the random phase approximation (RPA).
Several bound states have emerged in the correlation function, which correspond 
well to the magnons and excitons in the RIXS experiment.
Thus the weak-coupling theory based on the itinerant electron picture 
could provide a good explanation of the excitation spectra, although
there remains an issue whether the system 
really behaves like the Mott insulator or the band insulator 
\cite{Arita2012,Moser2014}.

\begin{figure}
\includegraphics[width=8.0cm]{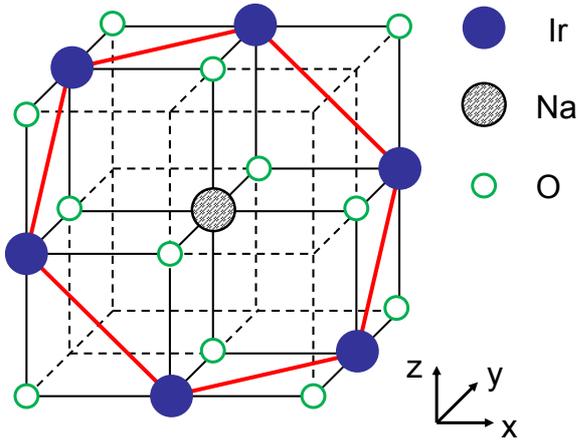}
\caption{\label{fig.structure}
Crystal structure of Na$_2$IrO$_3$ in the cubic setting.}
\end{figure}

\begin{figure}
\includegraphics[width=8.0cm]{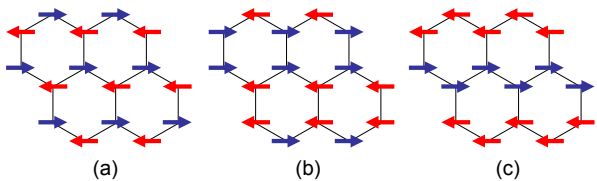}
\caption{\label{fig.pattern}
Possible magnetic orders in the honeycomb lattice;
(a)the N\'{e}el order, (b)the stripy order, and 
(c)the zigzag order, respectively.}
\end{figure}

Another fascinating example is Na$_2$IrO$_3$, which we will study in this
paper. 
It crystallizes in the space group $C2/m$ \cite{Choi2012,Ye2012,Lovesey2012},
where Ir$^{4+}$ ions constitute approximately honeycomb layer 
with a Na ion located at its centre as shown in figure \ref{fig.structure}.
It is an insulator with a temperature independent optical gap
$\sim 350$ meV \cite{Comin2012}.
Although the exotic spin liquid ground state had been expected originally,
it is found the magnetic order sets in below $T_{\textrm{N}}=15$ K
\cite{Singh2010}.
The type of the magnetic order is determined as a zigzag spin 
alignment shown 
in figure \ref{fig.pattern}(c) \cite{Choi2012,Ye2012,Comin2012,Liu2011}.
The low-lying excitations have been detected by inelastic neutron
scattering (INS) \cite{Choi2012} and RIXS \cite{Gretarsson2013-1,
Gretarsson2013-2},
in which the magnetic and excitonic excitations are assigned.

Several theoretical studies have already been carried out to explain those 
characteristics. The Kitaev-Heisenberg spin model 
\cite{Kitaev2006,Chaloupka2010,Kimchi2011,Chaloupka2013,Sizyuk2014}
and a generic spin model \cite{Katukuri2014,Rau2014}
have been derived by the strong coupling expansion based on the localized 
electron picture 
and the phase diagram has been examined in a wide range of parameter space. 
The zigzag alignment, unfortunately, seems to be realized only in rather 
extreme parameter values. 
Recently a spin model containing spin-spin anisotropic exchange couplings
has been derived from the \emph{ab initio} calculation, 
having led to the zigzag 
order in the ground state \cite{Yamaji2014}.
As for the itinerant electron picture, the band calculations based on the 
DFT have been carried out \cite{Mazin2012,Foyevtsova2013,Kim2014}. 
It is found that quasimolecular orbitals are formed within a hexagon of six 
Ir ions, resulting in the insulating ground state.
Recently the self-interaction correction (SIC) has been done 
to the DFT \cite{Kim2014}, and makes the zigzag spin order stabilize 
in the ground state. 
Accordingly the itinerant electron picture may provide a good starting point.

When we turn our attention to the collective excitation,
three characters of the intensity of the excitation
spectrum are identified by the experiment \cite{Choi2012,
Gretarsson2013-1,Gretarsson2013-2}
:magnetic excitation for $\omega < 40 $ meV,
excitonic excitation around $\omega \sim$ 450 meV, and contribution
from the continuum part 600 meV $< \omega$.
Note that the last contribution is regarded as the transfer between
$j_{\textrm{eff}}=\frac{3}{2}$ and $\frac{1}{2}$ in the 
localized electron picture.
Despite intensive theoretical effort, collective excitations of Na$_2$IrO$_3$ 
have not been studied yet on the itinerant electron picture.
In this paper, we investigate the collective excitations with the weak-coupling theory based on the itinerant picture to address the excitation 
spectra.
To this end, we utilize a rather simple tight-binding model instead 
of \emph{ab initio}
calculation to clarify the underlying mechanism of the low-energy
excitations.  
Employing the Hartree-Fock approximation (HFA) to the tight-binding model, 
we first calculate the one-electron energy as well as the ground-state energy.
Then, we evaluate the density-density correlation function within the RPA 
with the help of the two-particle Green's function in order to investigate
the excitation spectra.

We consider the electron transfer mediated through O $2p$ orbitals
and the direct $d$-$d$ transfer between Ir ions.
When the direct $d$-$d$ transfer is weak, nearly flat and doubly
degenerate bands come out due to the formation of quasimolecular orbitals
as pointed out in previous studies \cite{Mazin2012,Foyevtsova2013,Kim2014}. 
Since one hole occupies in the $t_{2g}$ states per Ir ion and
two Ir atoms are contained in the unit cell of the honeycomb lattice,
the uppermost band is fully empty, thereby resulting in the
energy gap between the occupied and unoccupied energy bands.
The Coulomb interaction and magnetic order have minor influence on the
formation of the energy gap, and hence the system may be called a band 
insulator. 
The zigzag magnetic order is found the most stable when the direct $d$-$d$ 
transfer is weaker than a certain value. 
However, the energy differences between the zigzag spin state
and others such as the stripy, the N\'{e}el states
are as small as several meV's per Ir ion. 
Then, we calculate the density-density correlation function within the RPA
on the zigzag order in ground state. 
In our calculation, the collective excitations come out as bound states 
as well as quasi-bound
states with modifying the individual excitations of the electron-hole pair
creation. 
With an appropriate value of the direct $d$-$d$ transfer,
we find several bound states with the excitation
energy concentrated on $\omega < 50$ meV, which may be assigned as magnetic excitations.
Other several bound states emerge in the energy region
between $\omega \sim 350$ meV
and the bottom of the energy continuum,
which may be assigned as excitonic excitations.
These features of excitation spectra semi-quantitatively capture the 
characteristic of the observed results mentioned above.

With increasing magnitude of the direct $d$-$d$ transfer between Ir ions, 
the nearly flat bands get dispersive, 
but the system remains still insulating state even for the sizable
direct transfer. Within the HFA, the most stable magnetic order is, 
however, changed from the zigzag order to the N\'{e}el order 
when the direct $d$-$d$ transfer exceeds a certain value.
In such a situation when the direct $d$-$d$ transfer is substantial,
within the RPA on the N\'{e}el order, we have the bound states 
in the density-density correlation function, 
which are distributed in rather wide range of energy.
Since this behavior is quite different from the experiment,
the large direct $d$-$d$ transfer is unfavourable in real materials.
Actually a recent analysis of the tight-binding parameters by the 
\emph{ab initio} calculation has claimed that the structural distortions 
of all types suppress the direct $d$-$d$ transfer \cite{Foyevtsova2013}.

The present paper is organized as follows. 
In \S\ref{sect.2},
we introduce a multi-orbital tight-binding model.
In \S\ref{sect.3}, the electronic structure is evaluated 
within the HFA. 
In \S \ref{sect.4}, we calculate the density-density 
correlation function within the RPA.
The excitation spectra are evaluated at the special $\textbf{k}$ spots.
Section \ref{sect.5} is devoted to the concluding remarks. 

\section{Model Hamiltonian\label{sect.2}}
Similar to the case of Sr$_2$IrO$_4$, each Ir ion in Na$_2$IrO$_3$ 
resides around the
centre of oxygen octahedra.
Due to the crystal electric field of IrO$_6$,
the energy level of the $e_g$ orbitals of Ir atom is
about 2-3 eV higher 
than that of the $t_{2g}$ orbitals. We take account of only $t_{2g}$ orbitals,
and ignore the small lift of degeneracy arising from the 
distortion of IrO$_6$ octahedra in a first step. 
Thereby, as a minimal model, the Hamiltonian of a multi-orbital Hubbard model
is defined on a honeycomb lattice,
\begin{equation}
 H = H_{\rm SO}+H_{\rm I}+ H_{\rm kin},
\end{equation}
with
\begin{eqnarray}
H_{\rm SO} & = & \zeta_{\rm SO}\sum_{i}\sum_{nn'\sigma\sigma'}
 d_{in\sigma}^{\dagger}({\bf L})_{nn'}
 \cdot({\bf S})_{\sigma\sigma'}d_{in'\sigma'}, \\
H_{\rm I} & = & 
  U\sum_{i,n} n_{in\uparrow}n_{in\downarrow} \nonumber \\
  &+&
 \sum_{i,n<n'\sigma}[U' n_{in\sigma}n_{in'-\sigma}
                 + (U'-J) n_{in\sigma}n_{in'\sigma}] \nonumber\\
 &+&J\sum_{i,n\neq n'} (d_{in\uparrow}^{\dagger}d_{in'\downarrow}^{\dagger}
                     d_{in\downarrow}d_{in'\uparrow}
                    +d_{in\uparrow}^{\dagger}d_{in\downarrow}^{\dagger}
                     d_{in'\downarrow}d_{in'\uparrow}), \nonumber\\
\label{eq.H_Coulomb} \\
H_{\rm kin} & = & \sum_{\left\langle i,i'\right\rangle }
\sum_{n,n'\sigma}[\hat{T}_{i,i'}]_{n,n'}d_{in\sigma}^{\dagger}d_{i'n'\sigma}
+ {\rm H.c.},
\end{eqnarray}
where $H_{\rm SO}$, $H_{\rm I}$, and $H_{\rm kin}$ are described by 
the annihilation ($d_{in\sigma}$) and creation 
($d_{in\sigma}^{\dagger}$) operators of $5d$ electron with orbital 
$n$ ($=yz,zx,xy$) and spin $\sigma$ at the Ir site $i$.
The $\langle i,i'\rangle$ indicates the nearest neighbour sum, and
$n_{in\sigma}\equiv d_{in\sigma}^{\dagger}d_{in\sigma}$.

The $H_{\rm SO}$ describes the SOI of $5d$ electrons where 
${\bf L}$ and ${\bf S}$ denote the orbital and spin angular momentum 
operators, respectively. 
We use the value $\zeta_{\rm SO}=0.4$ eV in the following
calculation. The $H_{\rm I}$ represents the Coulomb interaction
between the $t_{2g}$ electrons.
Parameters satisfy $U=U'+2J$ \cite{Kanamori1963}.
We use the values $U=1.4$ eV, and $J/U=0.15$ in the following calculation. 
These parameter values for Ir atom have been utilized also in
Sr$_2$IrO$_4$ \cite{Igarashi2013-1,Igarashi2013-2}.

The $H_{\rm kin}$ stands for the kinetic energy described by the 
hopping matrix $\hat{T}_{i,i'}$.
For simplicity, only transfers between the nearest neighbour
Ir ions are taken into account. 
There are three types of bond between the nearest neighbour 
Ir ions, and we call them as bond 1, 2, and 3 as illustrated in
figure \ref{fig.Bz} (a).
Then, two kinds of electron transfer may contribute to
$\hat{T}_{i,i'}$ between the adjacent Ir-Ir pair.
The one is indirect transfer 
via oxygen $2p$-orbitals ($\hat{T}_{i,i'}^{(pd)}$), 
and the other is direct
transfer between the Ir $5d$ orbitals ($\hat{T}_{i,i'}^{(dd)}$).
The former could
take place only between the different $5d$ orbitals. It may be 
expressed in a matrix form with the bases $n=yz$, $zx$, $xy$ in order:
\begin{equation}
 \hat{T}_{i,i'}^{(pd)} 
= 
\left(\begin{array}{ccc}
                        0 & 0 & -t_p \\
                        0 & 0 &  0 \\
                       -t_p & 0 &  0 
                          \end{array} \right), 
\left(\begin{array}{ccc}
                        0 & 0 &  0 \\
                        0 & 0 &  t_p \\
                        0 & t_p &  0 
                          \end{array} \right), 
                  \left(\begin{array}{ccc}
                        0 & t_p &  0 \\
                        t_p & 0 &  0 \\
                        0 & 0 &  0 
                          \end{array} \right),
\end{equation} 
for $\langle i,i'\rangle$ belonging to 
bonds 1, 2, and 3, respectively.
Here $t_p$ may be evaluated by 
\begin{equation}
   t_p =V_{pd\pi}^2/E_{pd}, 
\end{equation}
where $V_{pd\pi}$($=V_{pd\sigma}/\sqrt{3}$) stands for the Slater-Koster 
mixing parameter \cite{Slater1954} between the O $2p$ and Ir $5d$ orbitals, 
and $E_{pd}$ 
denotes
the charge-transfer energy from Ir $5d$ orbitals to O $2p$ orbitals. 
According to Harrison's procedure \cite{Harrison}, 
$V_{pd\sigma}$ and $E_{pd}$ may be 
estimated as $-1.8$ eV and $3$ eV, respectively, and hence
$t_p$ is estimated as $0.375$ eV. 
As regards the direct $d$-$d$ transfer, the matrix
$\hat{T}_{i,i'}^{(dd)}$ may be expressed by means of the conventional
Slater-Koster parameters $V_{dd\sigma}$, $V_{dd\pi}$ 
($=-2V_{dd\sigma}/3$),
and $V_{dd\delta}$ ($=V_{dd\sigma}/6$). 
Notice that 
it has non-zero matrix elements between the same $5d$ orbitals as well as
between the different $5d$ orbitals.
With the help of Harrison's procedure,
$V_{dd\sigma}$ is estimated as $-0.51$ eV. 
Since there exist several complications in real crystals,
these values of transfer are regarded as providing only order of magnitude.

\section{Electronic structure within the Hartree-Fock 
Approximation\label{sect.3}}
\begin{figure}
\includegraphics[width=8.0cm]{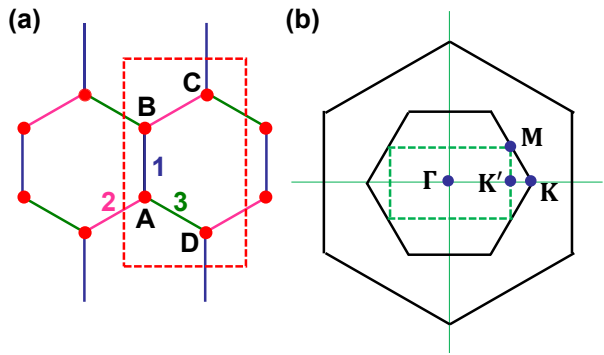}
\caption{\label{fig.Bz} 
(a) Honeycomb lattice. Three types of bond represented by
blue, magenta, and green lines are labeled as 1, 2, and 3, respectively. 
The unit cell consists of ions A and B.
In the magnetic ordering phases, Ir atoms A, B, C, and D 
form a unit cell, which is enclosed by a broken line.
(b)Reciprocal lattice and corresponding Brillouin zones.
The hexagon attached by K and M forms the first Brillouin zone 
in the absence of magnetic orders, 
while the rectangle enclosed by a broken line forms 
the magnetic Brillouin zone.
Centre and corners of larger hexagon represent the reciprocal lattice points.
}
\end{figure}

Before going to the HFA, it is instructive to investigate
the situation with no
Coulomb interaction working. Without magnetic orders, we can define
a unit cell containing Ir ions A and B shown in  figure \ref{fig.Bz}(a). 
The corresponding first Brillouin zone (BZ) is illustrated as a smaller 
hexagon in figure \ref{fig.Bz}(b).
The one-electron energy is evaluated by diagonalizing 
$H_{\rm kin}+H_{\rm SO}$ for momenta in the BZ.

\begin{figure}
\includegraphics[width=8.0cm]{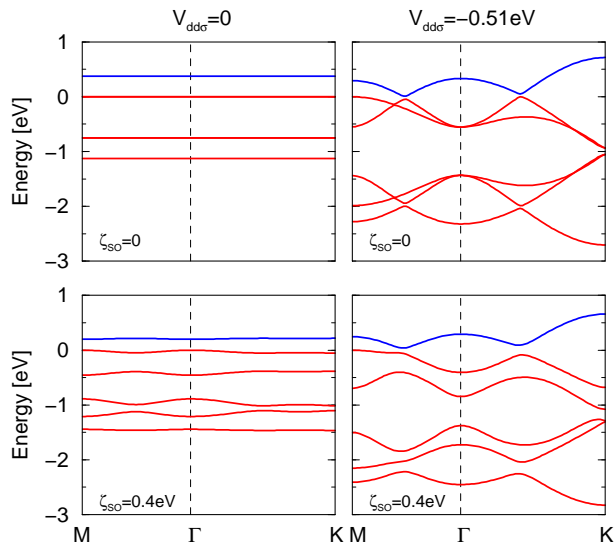}
\caption{\label{fig.hexagon} 
One-electron energy as a function of momenta along symmetry lines,
when the Coulomb interaction is disregarded.
The origin of energy is set at the top of the valence band.
The conduction band is doubly degenerated.
The left (right) panels show
the results in the absence (presence) of direct $d$-$d$ transfer.
The top (bottom) panels illustrate
the results in the absence (presence) of the SOI.}
\end{figure}

Figure \ref{fig.hexagon} shows the one-electron energy with momenta along 
symmetry lines.  
The origin of the energy is set at the top of the valence band. 
Each line is doubly degenerated. 
In the absence of the direct $d$-$d$ transfer, the dispersion is flat 
or nearly flat for $\zeta_{SO}=0$ or $\zeta_{SO}=0.4$ eV, respectively 
(left top or bottom panel of figure \ref{fig.hexagon}).
It indicates the formation of molecular orbitals within the Ir hexagon.
Since one hole exists per Ir site and two Ir ions are contained in a unit cell, 
the uppermost band should be empty, indicating a non-magnetic insulating 
ground state. Then, introduction of the direct $d$-$d$ transfer changes 
the situation. As shown in the right panels of figure \ref{fig.hexagon}, 
it makes the bands dispersive, but the system is narrowly insulating.  
This situation contrasts with that of Sr$_2$IrO$_4$, 
where the so-called ``$j_{\rm eff}=1/2$'' band, 
which is doubly degenerated when ignoring the Coulomb interaction, 
is half occupied, indicating that the system is a metal 
in the absence of the Coulomb interaction.

Now we consider the situation that the Coulomb interaction turns on,
and that a certain magnetic order such as the N\'{e}el, the stripy, or 
the zigzag orders, materializes in the ground state
(see figure \ref{fig.pattern}) \cite{Liu2011}. To describe these orders,
four sublattices A, B, C, and D, as shown in figure \ref{fig.Bz}(a),
are introduced.
The corresponding first magnetic Brillouin zone (MBZ) is denoted as the 
rectangle enclosed by the broken line in figure \ref{fig.Bz}(b).
With wave vector ${\bf k}$ in the first MBZ,
we define the Fourier transform of annihilation operator as 
\begin{equation}
 d_{\lambda n\sigma}({\bf k}) = (4/N)^{\frac{1}{2}}\sum_{j}d_{jn\sigma}
                             \textrm{e}^{-\\textrm{i}{\bf k}\cdot{\bf r}_j} ,
\label{eq.Fourier}
\end{equation}
where $j$ runs over the lattice sites belonging to one of the
four sublattices A, B, C, and D specified by $\lambda$,
and $N$ is the number of Ir ions.
 
With this notation, the one-electron energy 
$H_0\equiv H_{\rm kin}+H_{\rm SO}$ is rewritten as
\begin{equation}
 H_{0} = \sum_{{\bf k}\xi\xi'} d_{\xi}^{\dagger}({\bf k})
             \left[\hat{H}_0({\bf k})\right]_{\xi,\xi'}d_{\xi'}({\bf k}),
\label{eq.h0k}
\end{equation}
where use has been made of abbreviations 
$\xi=(\lambda,n,\sigma)$ and $\xi'=(\lambda',n',\sigma')$.
The $[\hat{H}_0({\bf k})]_{\xi,\xi'}$ denotes the matrix element of
the Fourier transform of $H_0$ expressed in the $\xi$ basis.

We carry out the HFA by following the conventional procedure as explained in 
\cite{Igarashi2013-1}. 
First, we rewrite the Coulomb interaction
as $H_{\rm I}$ $=$ $\frac{1}{2}\sum_{i}\sum_{\nu_1,\nu_2,\nu_3,\nu_4}
 g(\nu_1\nu_2;\nu_3\nu_4) d_{i\nu_1}^{\dagger}d_{i\nu_2}^{\dagger}
 d_{i\nu_4}d_{i\nu_3}$
where $\nu_{m} \equiv (n_{m},\sigma_{m})$ with $m=$ 1, 2, 3, 
and 4. By comparing this with (\ref{eq.H_Coulomb}), we can
determine the content of $g(\nu_1\nu_2;\nu_3\nu_4)$.
Then, we replace $H_{\rm I}$ by
\begin{equation}
 H_{\rm I}^{\rm HF} = \frac{1}{2}\sum_{j}\sum_{\xi_1,\xi_2,\xi_3,\xi_4}
 \Gamma^{(0)}(\xi_1\xi_2;\xi_3\xi_4)   
  \langle d_{j\xi_2}^{\dagger}d_{j\xi_4}\rangle d_{j\xi_1}^{\dagger}d_{j\xi_3},
\end{equation}
with
\begin{equation}
 \Gamma^{(0)}(\xi_1\xi_2;\xi_3\xi_4)=
  g(\xi_1\xi_2;\xi_3\xi_4)-g(\xi_1\xi_2;\xi_4\xi_3),
\end{equation}
where $\xi=(\lambda,\nu)$.
The $\langle X \rangle$ denotes the ground-state 
average of operator $X$.

The expectation values of the electron density contained in 
$H_{\rm I}^{\rm HF}$ has to be self-consistently determined.
For this purpose and evaluating the single-particle energy, 
it is convenient to introduce the single-particle Green's function 
in a matrix form with $24\times 24$ dimensions,
\begin{equation}
\left[\hat{G}({\bf k},\omega)\right]_{\xi,\xi'}=- \textrm{i} \int\langle 
 T[d_{\xi}({\bf k},t)d_{\xi'}^{\dagger}({\bf k},0)]\rangle
 {\rm e}^{\textrm{i} \omega t}{\rm d}t,
\label{eq.dG}
\end{equation}
where $T$ is the time ordering operator, and 
$X(t)\equiv {\rm e}^{\textrm{i} H't}
X{\rm e}^{-\textbf{i} H't}$ with $H'=H_{0}+H_{\rm I}^{\rm HF}$.
The Green's function can be solved
by diagonalizing the Hamiltonian matrix with $24\times 24$ dimensions.
Let the $\ell$-th energy eigenvalue for ${\bf k}$ be $E_{\ell}({\bf k})$,
and the corresponding wave function be $[\hat{U}({\bf k})]_{\xi,\ell}$.
Then the Green's function may be expressed as
\begin{equation}
[\hat{G}({\bf k},\omega)]_{\xi,\xi '}
=\sum_{\ell}
\frac{[\hat{U}({\bf k})]_{\xi,\ell}[\hat{U}({\bf k})^{-1}]_{\ell,\xi '}}
{\omega-E_{\ell}({\bf k})\pm \textbf{i} \delta}, 
\label{eq.Green}
\end{equation}
where $\delta$ denotes a positive convergent factor, and +(-) is taken when 
the energy level with $E_{\ell}({\bf k})$ is unoccupied (occupied).
Note that 
\begin{equation}
\langle d_{j\xi}^{\dagger}d_{j\xi'}\rangle=
\frac{4}{N}\sum_{{\bf k}}
\int [-\textbf{i} \hat{G}({\bf k},\omega)]_{\xi,\xi'}
{\rm e}^{\textbf{i}\omega0^{+}}\frac{{\rm d}\omega}{2\pi}. 
\label{eq.gap}
\end{equation}

Once we obtain the stable self-consistent solution, 
we could calculate the ground-state energy. Noting that
\begin{equation}
 \sum_{{\bf k}\ell} E_{\ell}({\bf k}) 
= \langle H_0\rangle + 2\langle H_{\textrm{I}}^{\textrm{HF}}\rangle,
\end{equation}
where the sum over $({\bf k}\ell)$ is restricted within the occupied levels,
we express the ground-state energy as
\begin{equation}
 \langle H'\rangle 
=\frac{1}{2}
\langle H_0 \rangle + \frac{1}{2}\sum_{{\bf k}\ell}E_{\ell}({\bf k} ).
\end{equation}
Here $\langle H_0 \rangle$ is evaluated by using (\ref{eq.h0k}) 
and (\ref{eq.Green}):
\begin{equation}
 \langle H_0\rangle =
 \sum_{{\bf k}\ell} \sum_{\xi\xi'} [\hat{U}({\bf k})^{-1}]_{\ell,\xi'}
  [H_0({\bf k})]_{\xi',\xi} [\hat{U}({\bf k})]_{\xi,\ell},
\label{eq.avH0}
\end{equation}
where the sum over $({\bf k}\ell)$ is again restricted within the occupied 
levels.

\subsection{Numerical calculation}
Since the staggered magnetic moment is directing along the crystal $a$ 
axis \cite{Ye2012,Liu2011}, we assume this is the direction of the 
staggered moment for the N\'{e}el, stripy, and zigzag orders 
in the self-consistent procedure. 
As already mentioned, the parameter values 
are set $\zeta_{SO}=0.4$ eV, $U=1.4$ eV, and $J/U=0.15$ in the following. 
As regards the transfer, we fix the strength of the indirect transfer 
by setting $V_{pd\sigma}=-1.84$ eV and $E_{pd}=3.01$ eV, or equivalently, 
$t_p=0.375$ eV. With evaluating (\ref{eq.avH0}), we carry out
the sum over ${\bf k}$ by dividing the MBZ into 
$120\times 60$ meshes. To achieve the convergence, 
the iteration of $100\sim 500$ times is necessary for some cases. 

\begin{figure}
\includegraphics[width=8.0cm]{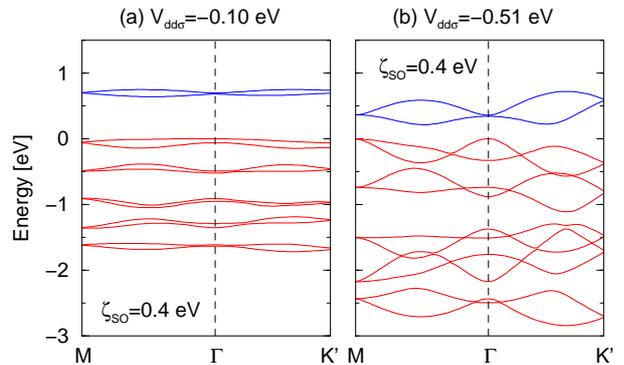}
\caption{\label{fig.disp} 
One-electron energy $E_{\ell}({\bf k})$ as a function of ${\bf k}$ along 
symmetry lines with  
(a)
$V_{dd\sigma}=-0.1$ eV in the zigzag phase
and (b) $V_{dd\sigma}=-0.51$ eV in the Ne\'{e}l phase.
$U=1.4$ eV and $U'/U=0.7$ with $2J+U'=U$.
Each band is doubly degenerated.
The origin of energy is set at the top of the valence band.}
\end{figure}

For $|V_{dd\sigma}|<0.23$ eV, we find the zigzag order
is the most stable one
among the zigzag, N\'{e}el, and stripy orders, 
but the energy differences
among those states are rather small. For $V_{dd\sigma}=-0.1$ eV, for instance,
the energy of the zigzag order is $0.007$ eV per Ir ion lower 
than that of the N\'{e}el order 
and $0.011$ eV lower than the energy of stripy order. 
The the orbital and spin moments are found to be parallel to each other
with $\langle L_a\rangle= \pm 0.265$ and $\langle S_a\rangle= \pm 0.085$,
hence the magnetic moment $0.435 \mu_{\textrm B}$. 
The latter value should be compared 
with the experimental value $0.22 \mu_{\textrm B}$.
Figure \ref{fig.disp}(a) shows $E_{\ell}({\bf k})$ as a function of 
${\bf k}$ along symmetry lines in the zigzag order for $V_{dd\sigma}=-0.1$ eV.
The dispersion curves experience weak dispersion, particularly
for the uppermost valence band and the conduction band. The band gap opens
with the size of $\sim 0.5-0.8$ eV. Note that it is
similar to the one-electron energy shown in the left panels of
figure \ref{fig.hexagon}. The latter comes out from the non-magnetic
state with the Coulomb interaction disregarded, implying that 
the energy gap between the occupied and unoccupied states 
is originated not from the Coulomb interaction but from the formation of 
quasimolecular orbitals on Ir hexagons \cite{Mazin2012,Foyevtsova2013}.
This interpretation is different from the one based on
the simple localized electron picture, which leads to the Kitaev-Heisenberg 
spin model in the strong coupling theory.
Note that the small dispersions of the conduction band and the uppermost 
valence band and the sizable band gap are comparable with 
those of the \emph{ab-initio} calculation 
(Figure 4 (b) in \cite{Kim2014} for example).

For $|V_{dd\sigma}|>0.23$ eV, we find the N\'{e}el state
is the most stable one. 
For, $V_{dd\sigma}= -0.51$ eV, which value may be estimated by the
Harrison's procedure,
the orbital and spin moments are found parallel
to each other with $\langle L_a\rangle= \pm 0.095$ and 
$\langle S_a\rangle= \pm 0.012$,
and hence the magnetic moment is $0.119 \mu_{\textrm B}$.
Figure \ref{fig.disp} (b) shows $E_{\ell}(\textbf{k})$ as a function
of $\textbf{k}$ for $V_{dd\sigma}=-0.51$ eV.
The dispersion depends considerably on ${\bf k}$ with rather small energy
gap. Since the zigzag order is confirmed in the real Na$_2$IrO$_3$, 
the small value of $|V_{dd\sigma}|$ is reasonable in this respect.\\

\section{\label{sect.4}Electron-hole pair excitations within the RPA}
To investigate the elementary excitations, we consider the density-density
correlation function defined by
\begin{equation}
\left[\hat{Y}^{+-}({\bf q},\omega)\right]_{\xi_1\xi'_{1};\xi\xi'}
= \int_{-\infty}^{\infty} 
\langle [\rho_{{\bf q}\xi_1\xi'_1}(t)]^{\dagger}
       \rho_{{\bf q}\xi\xi'}(0)\rangle {\rm e}^{\textbf{i}\omega t}{\rm d}t,
\end{equation}
with
\begin{equation}
  \rho_{{\bf q}\xi\xi'} 
= (4/N)^{\frac{1}{2}}\sum_{\bf k}
   d_{\xi}^{\dagger}({\bf k+q})d_{\xi'}({\bf k}).
\end{equation}
When ${\bf k+q}$ lies outside the first MBZ,
it is reduced back to the first MBZ by a reciprocal 
vector ${\bf G}$. 
Since $\hat{Y}^{+-}({\bf q},\omega)$ is a matrix of $576\times 576$ dimensions,
we consider a representative spectral function defined by
\begin{equation}
 I({\bf q},\omega) \equiv
 \sum_{\xi\xi'}\left[\hat{Y}^{+-}({\bf q},\omega)\right]_{\xi\xi';\xi\xi'}.
\end{equation}
To evaluate the correlation function, we introduce the time-ordered 
Green's function defined by
\begin{equation}
\left[\hat{Y}^{{\rm T}}({\bf q},\omega)\right]_{\xi_1\xi'_{1};\xi\xi'}
=-\textbf{i} \int \left\langle 
T\left\{[\rho_{{\bf q}\xi_1\xi'_{1}}(t)]^{\dagger}
\rho_{{\bf q}\xi\xi'}(0)\right\}\right\rangle
{\rm e}^{\textbf{i} \omega t}{\rm d}t,
\label{eq.green_time}
\end{equation}
and use the fluctuation-dissipation theorem 
for $\omega>0$ \cite{Igarashi2013-1},
\begin{equation}
 \left[\hat{Y}^{+-}(q)\right]_{\xi_1\xi'_{1};\xi\xi'}=
 -\textbf{i} \left\{\left[\hat{Y}^{{\rm T}}(q)
 \right]^{*}_{\xi\xi';\xi_1\xi'_{1}}
        -\left[\hat{Y}^{{\rm T}}(q)\right]_{\xi_1\xi'_{1};\xi\xi'}
\right\},
\label{eq.fdt1}
\end{equation}
where $q\equiv ({\bf q},\omega)$.
The Green's function $\hat{Y}^{+-}(q)$ can be evaluated by means of
the particle-hole propagator.
The derivation is concisely summarized below and
the detail is given in \cite{Igarashi2013-1}.

We take account of the multiple scattering between 
particle-hole pair within the RPA. 
Then, the Green's function is expressed as
\begin{equation}
\hat{Y}^{{\rm T}}(q)= \hat{F}(q)[\hat{I}+\hat{\Gamma}\hat{F}(q)]^{-1}
 = \left[\hat{F}(q)^{-1}+\hat{\Gamma}\right]^{-1},
\label{eq.time_ladder}
\end{equation}
where $\hat{I}$ is the unit matrix, and
\begin{equation}
[\hat{\Gamma}]_{\xi_{2}\xi'_{2};\xi_{1}\xi'_{1}}=
\Gamma^{(0)}(\xi_{2}\xi'_{1};\xi_{1}\xi'_{2}). 
\end{equation}
Here the particle-hole propagator $\hat{F}(q)$ is defined as
\begin{eqnarray}
&&
[\hat{F}(q)]_{\xi_2\xi'_{2};\xi_1\xi'_{1}} 
\label{eq.defF} \\
&\equiv&
-\textbf{i} \frac{4}{N}\sum_{{\bf k}}\int\frac{{\rm d}k_{0}}{2\pi}
[\hat{G}({\bf k+q},k_{0}+\omega)]_{\xi_{2}\xi_{1}}
[\hat{G}({\bf k},k_{0})]_{\xi'_{1}\xi'_{2}}.
\nonumber
\end{eqnarray}
\begin{widetext}
By substituting (\ref{eq.Green}) into the single-particle 
Green's function, and carrying out the integration over $k_0$ in 
(\ref{eq.defF}), we have
\begin{eqnarray}
[\hat{F}(q)]_{\xi_2\xi'_{2};\xi_1\xi'_{1}} 
 & = & \frac{4}{N}\sum_{{\bf k}}\sum_{\ell,\ell '}
 U_{\xi_{2}\ell}({\bf k+q})U_{\xi_{1}\ell}^{*}({\bf k+q})
 U_{\xi'_{1}\ell '}({\bf k})U_{\xi'_{2}\ell '}^{*}({\bf k})
\nonumber \\
 & \times & \left[\frac{[1-n_{\ell}({\bf k+q})]n_{\ell '}({\bf k})}
 {\omega-E_{\ell}({\bf k+q})+E_{\ell '}({\bf k})+\textbf{i} \delta}
 -\frac{n_{\ell}({\bf k+q})[1-n_{\ell '}({\bf k})]}
 {\omega-E_{\ell}({\bf k+q})+E_{\ell '}({\bf k})-\textbf{i} \delta}\right].
 \label{eq.green_pair}
\end{eqnarray}
\end{widetext}

Equation (\ref{eq.time_ladder}) contains
collective modes, which appear as bound states 
in the low-energy sector below the energy continuum of individual
electron-hole pair excitations. 
Since $\hat{F}(q)$ is an Hermite matrix, $\hat{F}(q)^{-1}+\hat{\Gamma}$ 
can be diagonalized by a unitary matrix.
If an eigenvalue becomes zero at $\omega=\omega_{\textrm{B}}({\bf q})$,
$\omega_{\textrm{B}}({\bf q})$ is regarded as the bound-state energy.
Let the corresponding eigenvector be $B_{\xi\xi'}({\bf q})$. Then, expanding 
$[\hat{Y}^{{\rm T}}(q)]_{\xi_1\xi'_{1};\xi\xi'}$ around 
$\omega=\omega_{\textrm{B}}({\bf q})$, we have 
\begin{equation}
\left[\hat{Y}^{{\rm T}}(q)\right]_{\xi_1\xi'_{1};\xi\xi'}=
\frac{[\hat{C}({\bf q})]_{\xi_1\xi'_{1};\xi\xi'}}
     {\omega-\omega_{\textrm{B}}({\bf q})+\textbf{i} \delta},
\label{eq.bound2}
\end{equation}
with 
\begin{eqnarray}
&&
[\hat{C}({\bf q})]_{\xi_1\xi'_{1};\xi\xi'}
\label{eq.wt.c} \\
&=&\frac{B_{\xi_1\xi'_{1}}({\bf q})
B_{\xi\xi'}^{*}({\bf q})}
{\sum_{\xi_2\xi'_{2}\xi_3\xi'_{3}}B_{\xi_3\xi'_{3}}^{*}({\bf q})
\frac{\partial [\hat{F}({\bf q},\omega_{\textrm{B}}({\bf q}))^{-1}]_{\xi_3\xi'_{3};\xi_2\xi'_{2}}}
     {\partial\omega}B_{\xi_2\xi'_{2}}({\bf q})}.
\nonumber
\end{eqnarray}
Inserting (\ref{eq.bound2}) into the right hand side of 
(\ref{eq.fdt1}), we have the correlation function,
\begin{equation}
 \hat{Y}^{+-}(q)=2\pi
\hat{C}({\bf q})\delta(\omega-\omega_{\textrm{B}}({\bf q})).
\label{eq.bound3}
\end{equation}

\subsection{Numerical calculation}
We evaluate $\hat{F}(q)$ by summing over 
${\bf k}$ in (\ref{eq.green_pair}) with dividing the first MBZ 
into $40\times 30$ meshes. 
The bound states are determined by searching for  
$\omega$ to give zero eigenvalue in $\hat{F}(q)^{-1}+\hat{\Gamma}$
within the accuracy of 0.001 eV.
The corresponding intensities are evaluated from finite difference 
between $\omega=\omega_{\textrm{B}}({\bf q})$ 
and $\omega=\omega_{\textrm{B}}({\bf q})+0.001$eV
in (\ref{eq.wt.c}) in place of the differentiation.
When $\omega$ enters into the energy continuum of individual electron-hole 
pair excitations, we need to evaluate the imaginary part arising from
the denominator in (\ref{eq.green_pair}). 
To make a rough estimate, we sort
each $E_{\ell}({\bf k+q})-E_{\ell'}({\bf k})$ inside
the energy continuum in (\ref{eq.green_pair}) into segments 
with the width of $0.05$ eV for $40\times 30$ ${\bf k}$-points, 
resulting in the histogram representation. Setting $\omega$ at the
centre of each segment, we evaluate (\ref{eq.green_pair}) and thereby
(\ref{eq.time_ladder}). 

\begin{figure} 
\includegraphics[width=8.0cm]{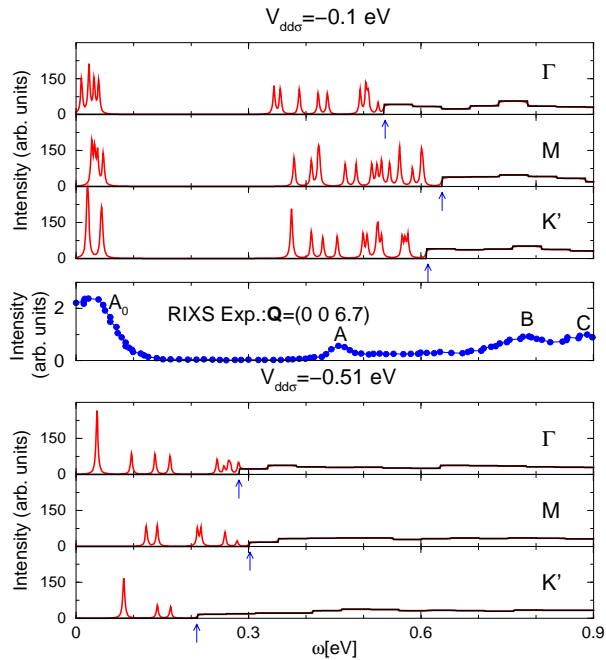}%
\caption{\label{fig.spectra}
Spectral function $I({\bf q},\omega)$ as a function of $\omega$  for  
${\bf q}$ at $\Gamma$, $M$, and $K'$ points; the top panel for  
$V_{dd\sigma}=-0.1$ eV and the bottom panel for $V_{dd\sigma}=-0.51$ eV.  
The middle panel is the RIXS spectra  
taken from \cite{Gretarsson2013-2} 
for ${\bf Q}=(0,0,6.7)$ in the $C2/m$ notation. 
The $\delta$-function peaks for bound states and the histograms for 
the continuum states are convoluted with the Lorentzian function of the 
half width half maximum 0.002 eV. 
The arrows indicate the lower boundary of the continuum spectra. 
} 
\end{figure}

For $|V_{dd\sigma}|<0.23$ eV,
the zigzag order becomes the ground state  
as already mentioned\cite{Com4}.
As a typical example in this region, we calculate the spectral  
function for $V_{dd\sigma}=-0.1$ eV. 
Top panel in figure \ref{fig.spectra} shows $I({\bf q},\omega)$ 
as a function of $\omega$ for ${\bf q}$ at $\Gamma$, $M$, and $K'$ points. 
We have four bound states clustered in the region where  
$\omega$ is less than 50 meV 
for all the ${\bf q}$-points.  
They may be called as \emph{magnetic} excitations ,which correspond
to the hump A$_0$ in the RIXS spectra.
Experimentally, INS detected a magnon mode below 6 meV \cite{Choi2012}  
while RIXS identified another around 35 meV \cite{Gretarsson2013-2}, 
which correspond well to our calculated results. 
The lowest excitation energy at the $\Gamma$ point is found 
around 9 meV. 
These magnetic excitations have quite a different origin 
from those of the Kitaev-Heisenberg spin model, 
since the former arise from the insulator based on the formation of 
quasimolecular orbitals on Ir hexagons, 
while the latter are brought about from the  model based on the localized  
electron picture in the strong coupling theory. 
There are no spectral intensities in the region between  
$0.050$ eV$<\omega<0.340$ eV. 
In addition, we have found about a dozen of bound states in the region  
between $\omega=$0.340 eV
and at the bottom of the energy continuum of individual 
electron-pair excitations, 
which may be called as the \emph{excitonic} excitations
and correspond to the RIXS spectra around the hump A. 
The continuous spectra start around $\omega=0.55-0.65$ eV,
which may correspond to the RIXS spectra around the humps
B and C. 
Accordingly, these capture qualitatively the characteristic 
of the RIXS spectra shown  
in the middle panel of figure \ref{fig.spectra}. 
 
For $|V_{dd\sigma}|>0.23$ eV, the N\'{e}el state becomes the ground state. 
As a typical example in this region, we calculate the spectra for  
$V_{dd\sigma}=-0.51$ eV, which is shown in the 
bottom panel of figure \ref{fig.spectra}. 
The bound states are distributed over a wide range of energy region  
below the continuum of electron-hole pair excitations.  
The number of the bound states are 9, 4, and 6 for ${\bf q}$ 
at $\Gamma$, $M$, and $K'$ points,  
respectively. The magnetic and the excitonic 
excitations are not sharply separated. Such spectral distribution is quite 
different from the RIXS spectra shown on the middle panel in figure  
\ref{fig.spectra}.

\section{\label{sect.5}Concluding remarks}
We have studied excitation spectra in Na$_2$IrO$_3$ on the basis of
the itinerant electron picture. 
We have employed a multi-orbital tight-binding model with the electron 
transfer mediated via the O $2p$ orbitals as well as the direct $d$-$d$ 
transfer,
considering only the $t_{2g}$ orbitals for Ir ions on a honeycomb lattice.
We have calculated the one-electron energy as well as the ground-state
energy within the HFA, and then the density-density correlation function
within the RPA. 
When the direct $d$-$d$ transfer is weak, it is found that
the zigzag order becomes the ground state, and that the energy bands
have small dispersions, probably due to the formation of quasimolecular 
orbitals.
The energy differences between the zigzag order and 
other orders have been, however, as small as several meV per Ir ion. 
We have obtained the collective excitations as
bound states in the density-density correlation function. Magnetic excitations
have been concentrated in the energy region less than 50 meV, 
while excitonic excitations exist around $\omega>350$ meV, in qualitative
agreement with the RIXS spectra.
When the direct $d$-$d$ transfer has exceeded a certain value, 
on the other hand,
the N\'{e}el order has become the ground state with the energy bands rather 
dispersive.
The collective excitations have distributed over a wide range of excitation
energy, which behavior is at variance with the RIXS spectra.

The $d$-$d$ transfer estimated by Harrison's procedure is larger than
the critical value. The above
findings may indicate that, in Na$_2$IrO$_3$, the direct $d$-$d$ 
transfer is suppressed by 
the structural distortions of all types. 
This interpretation has been pointed out by
a detailed analysis based on the \emph{ab initio} band calculation
\cite{Foyevtsova2013}.
Our present result lends support to this understanding 
from the point of view of the excitation spectra.
To address this issue in more quantitative way, 
it may be necessary to calculate the excitation 
spectra in more realistic models.
Finally electron correlations beyond the HFA and RPA may have important
effects on the excitation spectra.  
Studies along these directions are left in future.

\begin{acknowledgments}
We thank M. Takahashi for estimating the Slater-Koster parameters and for
invaluable discussions.
This work was partially supported by a Grant-in-Aid for Scientific Research
from the Ministry of Education, Culture, Sports, Science, and Technology,
Japan. 
\end{acknowledgments}

\bibliographystyle{apsrev} 
\bibliography{paper}

\begin{thebibliography}{45}
\bibitem[{Com({\natexlab{a}})}]{Com3}
\bibinfo{note}{There exist several bound states between the magnon modes and
  the continuous states, which have been missed in Ref. [20].}
\bibitem[{Com({\natexlab{b}})}]{Com4}
\bibinfo{note}{For $|V_{dd\sigma}|<0.03$ eV, there appears a soft mode at
  finite q value in the spectral function, implying the zigzag order is
  unstable for such magneto-orbital density wave.}
\end{thebibliography}

\end{document}